\documentclass[%
 reprint,
 amsmath,amssymb,
 aps,
]{revtex4-1}

\usepackage{graphicx}
\usepackage{dcolumn}
\usepackage{bm}
\usepackage{epstopdf}
\usepackage[utf8]{inputenc}
\usepackage[english]{babel}
\usepackage{color}
\usepackage{hyperref}
\begin{document}

\title{Baryogenesis of the Universe in cSMCS Model plus Iso-Doublet Vector Quark}

\author{Neda Darvishi} \email{neda.darvishi@fuw.edu.pl}
\affiliation{Faculty of Physics,
University of Warsaw, \\
Pasteura 5, 02-093 Warsaw, Poland}

\begin{abstract}
CP violation of the SM is insufficient to explain the baryon asymmetry in the universe and therefore an additional source of CP violation is needed. Here the extension of the SM by a neutral complex scalar singlet with a nonzero vacuum expectation value (cSMCS) plus a heavy  vector quark pair is considered. This model offers the spontaneous CP violation and proper description in the baryogenesis, it leads strong enough first-order electro-weak phase transition to suppress the baryon-violating sphaleron process.

 
\end{abstract}

\maketitle

\section{Introduction}
\label{Introduction}

As far as we know, our universe is dominated by matter. The baryon asymmetry of the universe (BAU) could be generated during the electroweak phase transition, as disused in numerous studies, e.g.\cite{ Anderson, Cawington, Dine, Bacharev, Espinosa, Shaposhnikov, Nelson, Cohen, Cormelli, MCDonald}, trying to identify the source of this asymmetry.  
According to the Sakharov conditions \cite{sak}, in order to generate the BAU it is required to have, first, the violation of the baryon number, second, the violation of C and  CP symmetries and third, the existence of a non equilibrium processes, see also the reference ~\cite{gold}. It has become apparent that the electroweak Standard Model (SM) is unable to account for the observed magnitude of the BAU because the amount of CP violation within the SM is not sufficient~\cite{Gavela:1994ds,Gavela:1994dt,ber}.  In the present work it has been assumed that the generation of BAU is provided by the model with a neutral complex scalar singlet $\chi$, which accompanies the SM-like Higgs doublet $\Phi$, and an iso-doublet vector quarks $V_L+V_R$ . This kind of extension of the SM was discussed in the literature with various motivations, can be found in eg.\cite{Branco:2003rt,Profumo:2007wc,Lebedev:2010zg,Espinosa:2011eu,Gabrielli:2013hma,marco,Kozaczuk,Jiang:2015cwa,AlexanderNunneley:2010nw,Barger:2008jx,Kozaczuk}. 

Here the potential with a softly broken global U(1) symmetry is considered, leading the model called cSMCS,  also see \cite{Darvishi:2016gvm,Krawczyk:2015xhl,Bonilla:2014xba}. The additional source of CP violation is provided by a neutral complex scalar singlet $\chi$ with non-zero expectation value. The issue of the CP violation due to a complex singlet with a complex expectation value has been previously discussed, see \cite{Darvishi:2016gvm,Krawczyk:2015xhl}.
In the presence of an iso-doublet vector quark and a complex singlet, the Yukawa Lagrangian acquires additional terms. While diagonalizing the quark mass matrix the whole Lagrangian will be modified with new terms which are functions of the time-dependent phase (CP violating phase ~\cite{McDonald}). The appearance of these terms leads to the generation of a baryon asymmetry.
\\
The content of this paper is as follows. In section \ref{smcs} a general presentation of the SMCS model and its constrained version (cSMCS) investigated in the paper is given. In particular, the subsection \ref{ext} describes the conditions for the spontaneous CP violation in the model. In section \ref{PHT}, the necessary conditions for strong enough first order electroweak phase transition in the model will be verified. 
The generation of a baryon asymmetry through the mixing of the SM quark and vector quarks is discussed in the section ~\ref{baryon}.
The section \ref{concl} contains our conclusion while detailed formula are presented in the Appendix.

\section{The cSMCS: The SM plus a complex singlet scalar}
\label{smcs}

The full Lagrangian of this model is given by
\begin{equation}
{ \cal L}={ \cal L}^{SM}_{ gf } +{ \cal L}_{scalar} + {\cal L}_Y(\psi_f,\Phi) + {\cal L}_Y(V_q,\chi), 
\label{lagrbas}
\end{equation}
where ${\cal L}^{SM}_{gf}$ describes the pure gauge boson terms as well the SM boson- SM fermion interaction, ${ \cal L}_{scalar}$ describes the scalar sector of the model with one SU(2) doublet $\Phi$ and a neutral complex scalar (spinless) singlet $\chi$. ${\cal L}_Y(\psi_f,\Phi)$ and ${\cal L}_Y(V_q,\chi)$ represent, respectively, the Yukawa interaction of $\Phi $ with SM fermions and the Yukawa interaction of singlet scalar with vector quarks. The neutral singlet $\chi$ does not couple to the SM fermions and therefore the singlet- SM fermion interaction is present only through the mixing of the singlet $\chi$ with the doublet $\Phi$ (the same holds for the singlet interaction with the gauge bosons). The SM-like Higgs boson in the model predominantly consists of a neutral CP-even component of the $\Phi$ doublet and its mass is $\sim 125 $ GeV. There are two other higgs particles-like, see discussion in \cite{Darvishi:2016gvm,Krawczyk:2015xhl,Bonilla:2014xba}. 
\\
\\
We assume $\Phi$ and $\chi$ fields have non-zero vacuum expectation values ($vev$) $v$ and $w e^{i\xi}$, respectively ($v,w,\xi\in \bf{R}$). We shall use the following field decomposition around the vacuum state,
\begin{equation}
\Phi = \left( \begin{array}{c} \phi^+ \\   {1\over \sqrt{2}}(v + \phi_1 + i \phi_4) \\ \end{array} \right), 
\chi = {1\over \sqrt{2}}( w e^{i \xi} + \phi_2 + i \phi_3), \label{dec_singlet}
\end{equation}
where
\begin{equation}
w e^{i\xi}={w\cos\xi}+iw\sin\xi=w_1+iw_2. 
\label{VEV}
\end{equation}
Masses of the gauge bosons are given by the $vev$ of the doublet, e.g $M_W^2 = g^2 v^2/4$ for the $W$ boson.
\subsection{Potential}
The scalar potential of the model can be written as follows  \cite{Darvishi:2016gvm,Krawczyk:2015xhl,Bonilla:2014xba}
\begin{equation}
V=V_{D}+V_S+V_{DS}, \label{potgen}
\end{equation}
with the pure doublet and the pure singlet parts (respectively $V_{D}$ and $V_{S}$) and the mixed term $V_{DS}$. The SM part of the potential represent by $V_{D}$, is equal to
\begin{equation}
	V_{D} = - \frac{1}{2}{m_{11}^2} \Phi^\dagger\Phi 
	+ \frac{1}{2}\lambda_1 \left(\Phi^\dagger\Phi\right)^2.
\label{potSM}
\end{equation}

The potential for a complex singlet is equal to,
\begin{eqnarray}
V_{S} =&& -\frac{1}{2}m_s^2 \chi^* \chi -\frac{1}{2}m_4^2 (\chi^{*2} + \chi^2)
\nonumber\\&&
+ \lambda_{s1} (\chi^*\chi)^2 + \lambda_{s2} (\chi^*\chi)(\chi^{*2} + \chi^2) + \lambda_{s3} (\chi^4 + \chi^{*4})
\nonumber\\&&
+ \kappa_1 (\chi + \chi^*) + \kappa_2 (\chi^3 + \chi^{*3}) + \kappa_3(\chi+ \chi^*)(\chi^*\chi). \nonumber\\
\label{potS}
\end{eqnarray}
The doublet-singlet interaction term is,
\begin{eqnarray}
V_{DS} = &&\Lambda_1(\Phi^\dagger\Phi)(\chi^* \chi) + \Lambda_2 (\Phi^\dagger\Phi)(\chi^{*2}+ \chi^2)
\nonumber\\&&
+ \kappa_4 (\Phi^\dagger\Phi) ( \chi +\chi^*). 
\end{eqnarray}
There are three quadratic ($m^2_i$), six dimensionless quartic ($\lambda_i, \Lambda_i$) and four dimensionful parameters $\kappa_{i},\;i=1,2,3,4$, describing linear ($\kappa_1$) and cubic terms ($\kappa_2,\kappa_3$) and $\kappa_4$. The linear term $\kappa_1$ can be removed by a translation of the singlet field. Both $V_{S}$ and $V_{DS}$ are symmetry under the $\chi \to \chi^*$ transformation.
To simplify the model, we apply a global U(1) symmetry \cite{Darvishi:2016gvm,Krawczyk:2015xhl,Bonilla:2014xba}.
\begin{equation}
	U(1): \;\; \Phi \to \Phi,\, \chi \to e^{i\alpha} \chi .\label{u1def}
	\end{equation}

However, a non-zero $vev$ of $\chi$ would lead in such case to a spontaneous breaking of this symmetry and an appearance of massless Nambu-Goldstone scalar particles, what is not acceptable. Keeping some U(1)
soft-breaking terms in the potential would solve this problem. In what follows, we shall consider a potential with a soft-breaking of U(1) symmetry, where the singlet cubic terms $\kappa_{2,3}$, $\kappa_4$ and the singlet quadratic term $m_4^2$ are kept.
Therefore, we are left with the U(1)-symmetric terms ($m_{11}^2, m_{s}^2, \lambda_{1}, \lambda_{s1}, \Lambda_{1}$) and the U(1)-soft-breaking terms ($m_{4}^2, \kappa_{2,3,4}$). Simplifying slightly the notation by using:{{ $ \lambda_s= \lambda_{s1}, \Lambda=\Lambda_1$}}, we get the potential in the following form
\begin{eqnarray} \label{potchi}
V =&& -\frac{1}{2}{m_{11}^2} \Phi^\dagger\Phi+ \frac{1}{2}\lambda_1\left(\Phi^\dagger\Phi\right)^2 + \Lambda(\Phi^\dagger\Phi)(\chi^* \chi)
\nonumber\\&&
- \frac{1}{2}{m_s^2} \chi^* \chi + \lambda_{s} (\chi^*\chi)^2 + \kappa_4 (\Phi^\dagger\Phi) ( \chi +\chi^*)
\nonumber\\&&
 - \frac{1}{2}{m_4^2} (\chi^{*2} + \chi^2)+ \kappa_2 (\chi^3 + \chi^{*3}) + \kappa_3 (\chi + \chi^*)(\chi^*\chi). \nonumber\\
\label{potVS}
\end{eqnarray}
Note that $V$ is symmetry under the $\chi \to \chi^*$ transformation and we take all parameters real. Therefor $V$ is explicitly  CP conserving. We shall call the model with this choice of parameters,  cSMCS \cite{Darvishi:2016gvm,Krawczyk:2015xhl}. 
Note, that this potential (\ref{potchi}) is similar to the potential with two real singlets, with an additional $Z_2$ symmetry for the one singlet field, considered in paper \cite {marco}. In that model, however, CP violation is not possible.
\\
\\
The extremum conditions lead to the following constraints, 
\begin{equation}
	-m_{11}^2+v^2 \lambda_1 +2\sqrt{2} w_1 \kappa_4+\Lambda w^2 =0, \label{min1}
\end{equation}
\begin{eqnarray}
	w_1 (-\mu_1^2 +v^2 \Lambda + 2 w^2\lambda_{s}) + \sqrt{2}[3(w_1^2 - w_2^2) \kappa_2 \nonumber\\
	+ (3 w_1^2 + w_2^2)\kappa_3] +v^2\sqrt{2} \kappa_4= 0 , \label{min2} 
\end{eqnarray}
\begin{equation}
	w_2[-\mu_2^2+ v^2 \Lambda + 2 w^2 \lambda_{s} + 2 \sqrt{2} w_1 (- 3\kappa_2 + \kappa_3)]=0,
\label{min3} 
\end{equation}

Where the parameters $\mu_{1}^2$ and $\mu_{2}^2$ defined as
$$\mu_{1}^2=m_{s}^2+2m_4^2, \quad \mu_{_2}^2=m_{s}^2-2m_4^2. \label{mu}$$ 

Various spontaneous symmetry breaking extrema are possible, among them with vanishing one or two of vacuum expectation parameters $v, w_1, w_2$. Here we concentrate on the case with $v, w_1$ and $w_2$ different from zero, allowing for a vacuum which violate CP. Note, that for vanishing phase $\xi=0$ ($w_2=0$) CP violation is not possible for our model \cite{Darvishi:2016gvm}.
\\
In order to have a stable minimum, the parameters of the potential need to satisfy the positivity conditions. Positivity conditions read as follows:
\begin{equation}
\begin{array}{l}
 \lambda_1, \,\, \lambda_{s} > 0, \quad
 \Lambda > - \sqrt{2 \lambda_1 \lambda_{s}}.\\[3mm]
 \end{array} \label{pos}
\end{equation}
In addition the unitarity condition reaches to limit on quartic terms $\lambda_1$,  $\lambda_{s1}$ and $\Lambda$, below $4\pi$.
\subsection{The CP violating vacuum} \label{ext}
Minding the equation (\ref{min1}), when neither $v$, $w_1$ nor $w_2 $ vanish, an important relation can be obtained
via subtracting equation (\ref{min3}) from the equation (\ref{min2}),
\begin{equation}
	- 8 m_4^2 \cos^2\xi + 6 R_2 \cos\xi(1+2\cos2\xi)+2 R_3 \cos\xi + 
	 R_4 = 0,
	\label{CP2}
\end{equation}
where 
$$ R_2=\sqrt{2}w\kappa_2, \,\,  R_3=\sqrt{2}w\kappa_3, \,\,  R_4= {2\sqrt{2}v^2 \kappa_4\over w}\cos\xi, $$
all of which are of $[mass]^2$ dimension. In addition we have 
$$ R_4={v^2 \over w^2}(m_{11}^2- v^2 \lambda_1- w^2 \Lambda). $$
For a particular case, i.e. $R_2 = 0$, the above equation transforms to,
\begin{equation}
-8 m_4^2 \cos^2\xi+2 R_3 \cos\xi + R_4 =0.
\label{CP3}
\end{equation}

 In Fig.\ref{2.pdf} the regions allowed by equations (\ref{min1}),(\ref{min2}) and (\ref{min3}) of parameters for a vaccum with $v, w_1, w_2\neq0$  is presented. Fig. {\ref{2.pdf}(a) shows the region of parameters $R_3$, $R_4$ and $\xi$ as given by the Eq. ({\ref{CP3}}), for fixed $m_4^2$ is shown.\\
 In Fig.\ref{2.pdf}(b) and (c) the allowed regions of parameters $(R_3, R_4)$, for $R_3 = 0$, and the allowed region of the parameters $(R_2,R_4)$, for $R_3 = 0$ are shown, respectively. These regions are in agreement with the Eq. (\ref{CP2}), for fixed $m_4^2$.\\
The various aspect of CP violation with $\kappa_4=0$, are discussed in \cite{Darvishi:2016gvm}.
\begin{figure*}[ht]
\includegraphics[width=0.9\textwidth]{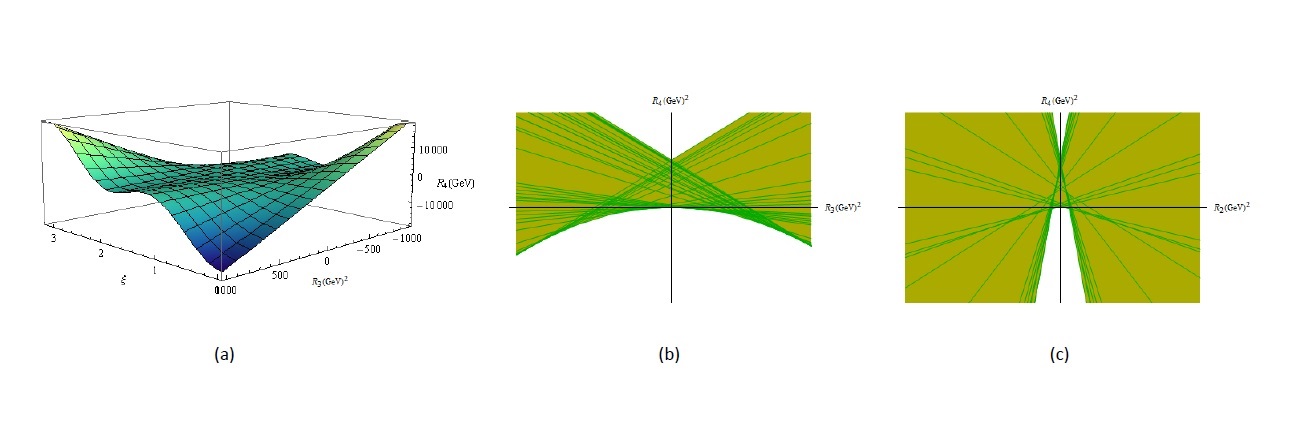}
\caption[CP violation]{{ {The spontaneously CP violation: the allowed (shaded) regions of the parameters $R_2$,$R_3$,$R_4$
for $-1 < \cos\xi < 1$ and $4 m_4^2=500\,$GeV$^2$. The boarder lines correspond to the $\cos \xi=\pm 1$ limits.
(a) Plot 3d for parameters $R_3$, $R_4$ and $\xi$ based on Eq. ({\ref{CP3}}); 
 (b)Regions for $R_3$ , $R_4$ allowed by Eq. (\ref{CP2}); 
(c) Regions for $R_2$ and $R_4$ given by Eq. (\ref{CP2}).}}}
\label{2.pdf}
\end{figure*}

\section{The electroweak phase transition} \label{PHT}
At very high temperatures far above the electroweak scale, the electroweak gauge symmetry is unbroken with no baryon number. The universe cools down and expands. Near the electroweak phase transition(EWPT) scale bubbles of broken electroweak symmetry appear and subsequently expand in the surrounding unbroken phase. Electroweak baryon asymmetry can be realized if this change of phase proceeds by a first-order phase transition.
If the phase transition is strongly first order, the baryon violating processes are out of equilibrium in the bubble walls and a net baryon number can be generated during the phase transition.
Phase transition is strong enough when \cite{Kuzmin,Kane,Quiros:1999jp},
 \begin{equation}
\frac{v({T_c})}{T_c}\geq 1,
\end{equation}
where $T_c$ corresponds to the critical temperature.\\
 To study the EWPT in the present model, we express the complex scalar $\chi$ in terms of its real and imaginary parts, $\chi=(\chi_1+i \chi_2)/\sqrt{2}$. For the potential at zero temperature we have
\begin{eqnarray}
 V(T_0)=&&-\frac{1}{2}{m_{11}^2} \Phi^\dagger\Phi+ \frac{1}{2}\lambda_1 \left(\Phi^\dagger\Phi\right)^2 
-\frac{{\mu}_1^2}{4}\chi_1^2
\nonumber\\&& 
-\frac{{\mu}_2^2}{4}\chi_2^2 +{1 \over 2}\Lambda (\Phi^\dagger\Phi) (\chi_1^2+\chi_2^2) 
	\nonumber\\&&
	+\frac{1}{4} \lambda_{s} (\chi_1^2+\chi_2^2)^2
  +\frac{1}{\sqrt{2}} \kappa_2 (\chi_1^3-3\chi_1\chi_2^2)
	\nonumber\\&&
	+\frac{1}{\sqrt{2}}\kappa_3(\chi_1^3+\chi_1\chi_2^2)+\sqrt{2}\kappa_4 (\Phi^\dagger\Phi) \chi_1.
 \label{potsS}
 \end{eqnarray}
The one-loop thermal corrections to the effective potential at finite temperature T are(see ref. ~\cite{Quiros:1999jp} for review),
\begin{equation}
\Delta V_{thermal} = \sum_{i} {n_i T^4 \over 2 \pi^2} I_{B,F} \left( {m_i^2 \over T^2} \right), 
\label{thermal}
\end{equation}
with
\begin{equation}
I_{B,F}(y) = \int_{0}^{\infty} dx \; x^2 \; ln \left[ 1 \mp e^{-\sqrt{x^2+y}} \right].
\label{I}
\end{equation}
The minus sign correspond to the bosons and the plus to sign for the fermions. In Eq.(\ref{thermal}) $m_i$ is the field-dependent mass and $n_i$ is the number of degrees of freedom, (see Appendix Eq.(\ref{A1})).
One-loop potential for our model, calculated using the high temperature approximation (i.e. keeping only $T^2$ terms) is as follows \cite{Dolan:1973qd},
\begin{eqnarray}
V(T)=&&\frac{1}{4}\overline{{m}}_{11}^2\phi_1^2+ \frac{1}{8}\lambda_1\phi_1^4
+ \frac{\overline{\mu}_1^2}{4}\phi_2^2+\frac{\overline{\mu}_2^2}{4}\phi_3^2\nonumber\\
 &&+\frac{1}{4}\Lambda \phi_1^2 (\phi_2^2+\phi_3^2) +\frac{1}{4}\lambda_{s}(\phi_2^2+\phi_3^2)^2\nonumber\\
 &&+ \kappa_2\frac{1}{\sqrt{2}}(\phi_2^3-3\phi_2\phi_3^2)+\kappa_3\frac{1}{\sqrt{2}}(\phi_2^3+\phi_2\phi_3^2)
     \nonumber\\
&&+\frac{1}{\sqrt{2}}\kappa_4 \phi_1^2 \phi_2+\overline{\kappa}_{34}{T^2\over 3} \phi_2,
 \label{potsS}
 \end{eqnarray}
where
\begin{eqnarray}
\overline{m}_{11}^2&& = -m_{11}^2+(3\lambda_1 + \Lambda+{2m_W^2+m_Z^2+2m_t^2 \over 2 v^2}){T^2\over 3},\nonumber \\
{1\over 2}\overline{\mu}_1^2&& = -{1\over 2}\mu_{1}^2 +(\Lambda+2\lambda_{s}){T^2\over 3}, \nonumber \\
{1\over 2} \overline{\mu}_2^2&& =-{1\over 2}\mu_{2}^2 +(\Lambda+2\lambda_{s}){T^2\over 3}, \nonumber \\
\overline{\kappa}_{34} &&=\sqrt{2}(\kappa_3+ \kappa_4).
\end{eqnarray}
The extremum conditions of the effective potential at temperature $T$, with respect to the fields $\phi_1,\phi_2$ and $\phi_3$, are
\begin{equation}
	\overline{m}_{11}^2+\lambda_1 v^2+ \Lambda w^2+2\sqrt{2} \kappa_4 w_1 =0, \\ \label{minT1}
\end{equation}
\begin{eqnarray}
	w_1 (\overline{\mu}_1^2 + \Lambda v^2 + 2 \lambda_{s} w^2) + \sqrt{2}\left[ 3 \kappa_2 (w_1^2 - w_2^2) \right. \nonumber\\
	+ \left. \kappa_3(3 w_1^2 + w_2^2) + \kappa_4 v^2 \right] + {2\over 3} \overline{\kappa}_{34} T^2 = 0, \nonumber\\
 \label{minT2} 
\end{eqnarray}
\begin{equation}
	w_2[\overline{\mu}_2^2+  \Lambda v^2 +2 \lambda_{s} w^2 + 2\sqrt{2} (- 3\kappa_2 + \kappa_3) w_1]=0.
\label{minT3} 
\end{equation}
At extreme temperatures, the solution of the equations (\ref{minT1}),(\ref{minT2}) and (\ref{minT3}) is
\begin{equation}
	v=0, \,\, w_1 \approx {\overline{\kappa}_{34} \over 2\lambda_s+\Lambda}, w_2\approx0 .
\end{equation}
Then, the scalar component, $\phi_3$, decouples from the model and therefore, the potential is similar to the SM plus a real singlet ~\cite{Branco:1998yk}.
\\
\\
Now, we will verify the parameter region where exists the strong first order phase transition, ${v(T_c)/ T_c}\geq 1$, with the critical temperature $T_c$ to be smaller than 250 GeV and $v(T_c)$ below its zero-temperature value $v_0 = 246$ GeV. We have performed a scan over the parameter space fulfilling unitarity and positivity conditions (see \cite{Darvishi:2016gvm}), namely: 
\begin{eqnarray}
-0.25 < \Lambda <0.25, \; 0 < \lambda_{s} < 1, && \; -1 <\rho_{2,3,4} < 1, \; 0 <\xi < \pi, \nonumber\\
 -90000 \,\, {\rm GeV}^2 < \mu_{1}^2, &&\, \mu_{2}^2,\, m_{11}^2 < 90000 \,\, {\rm GeV}^2,
\label{mml} 
\end{eqnarray}
where we used dimensionless parameters $\rho_{2,3,4}=\kappa_{2,3,4}/ w$.
Considering the SM-like scenarios at the LHC, the mass of lightest higgs boson in this model is given by $M_{h_1}^2 \approx m_{11}^2 \approx \lambda_1 v^2$ ($M_{h_1}$ $\approx$ 125 GeV). Thus, we take $\lambda_1$ in the range ~\cite{Darvishi:2016gvm}:
\begin{equation}
 0.2 < \lambda_{1} < 0.3. \label{l1}
\end{equation}
The model contains two additional higgs scalars $M_{h_2}$ and $M_{h_3}$, which we take to be ~\cite{Bonilla:2014xba,Darvishi:2016gvm}
\begin{equation}
M_{h_3} 
\gtrsim M_{h_2} > 150 \, {\rm GeV}.
\end{equation}
In~\cite{Darvishi:2016gvm} we have shown that these ranges of parameters are in agreement with LHC data and measurements of the oblique parameters.
\\
The results of our scan are shown in the Fig.\ref{EWPT}. In Fig.\ref{EWPT}(a) the allowed region of $v(T_c)/ T_c$ as a function of $T_c$ is shown. Within the interval $100<T_c<200$ the ratio $v(T_c)/ T_c$ ratio can reach $2.5$. The Fig.\ref{EWPT}(b) shows that $v(T_c)/ T_c \geq 1$ is possible for $|\rho_3| > 10^{-3}$.
\\
\\
We see that the strongly first-ordered EWPT is possible in our model. Since the out of equilibrium condition can be achieved for strong enough first order phase transition, in the bubble walls, we conclude that a successful BAU in our model is possible ~\cite{Quiros:1999jp}.
\begin{figure*}[ht]
\includegraphics[scale=.3]{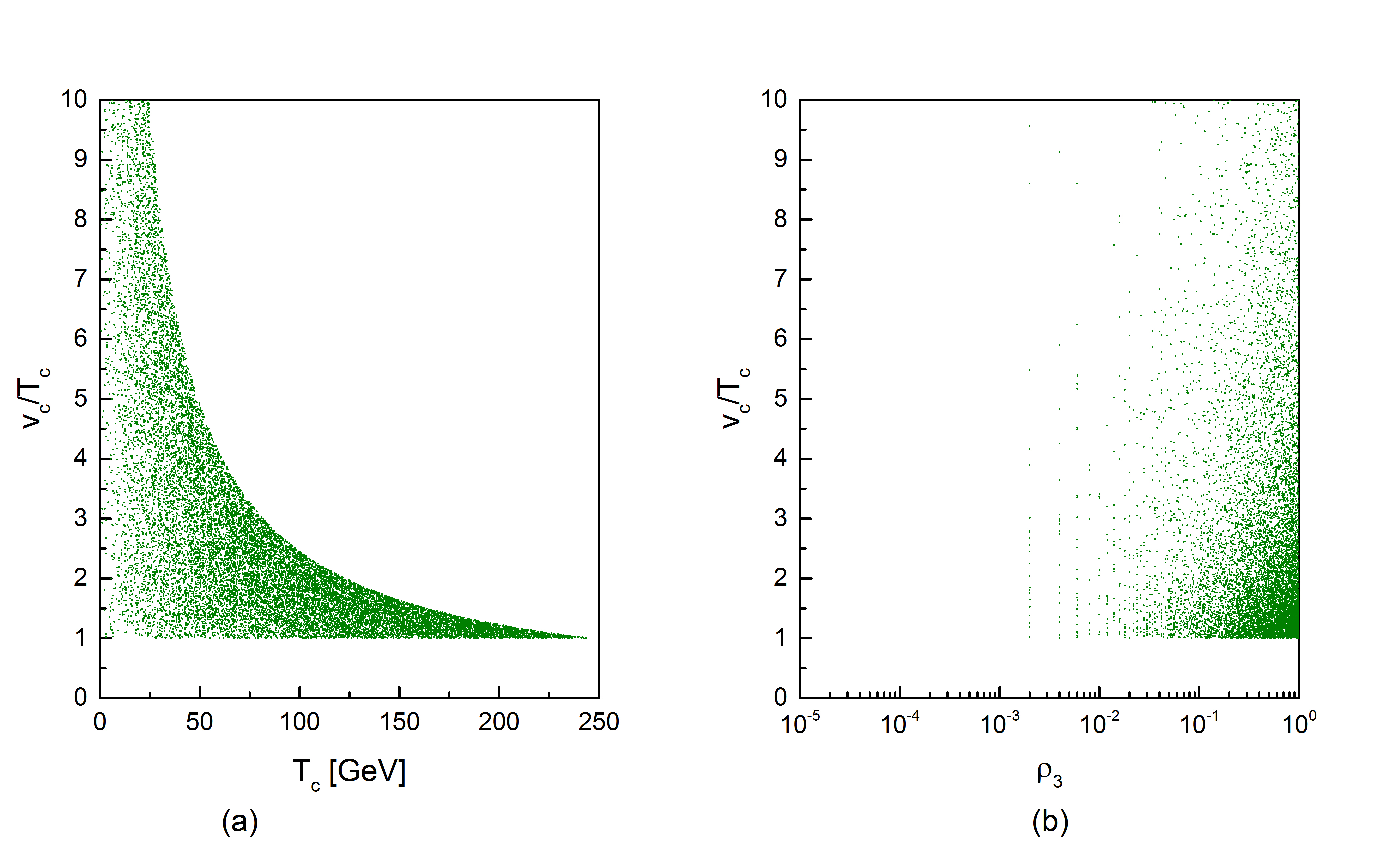}
\caption{The allowed regions of critical temperature $T_c$, $v_c \equiv v(T_c)$ and $|\rho_3|$  for strongly first order phase transition. (a) ($T_c$,$v_c/ T_c$) and (b) ($|\rho_3|$, $v_c/ T_c $). The scatter points are selected to satisfy the criterion, $(v(T_c)/ T_c)\geq 1$(see text for details).}
\label{EWPT}
\end{figure*}
\section{Baryogenesis}
\label{baryon}
In this section we describe the baryon asymmetry resulting from a mixing of the SM quarks and heavy vector quarks. 
We consider a pair of heavy iso-doublet vector quarks, $V_L+V_R$, with $V_L$ and $V_R$ having the same transformation properties under the gauge group of the SM as quark
doublet $Q_L$ ~\cite{Branco:1998yk,McDonald}.
The mass terms in the presence of the complex singlet are ( see Eq.(\ref{lagrbas})):
\begin{equation}
 {\cal L}_Y(V_{q},\chi)=\lambda_V \chi \overline{Q}_L V_R + M \overline{V}_L V_R+ h.c,
\label{LY}
\end{equation}
Here, we consider only one (the heaviest) quark doublet  $Q_L$ .
 To generate baryon asymmetry, the phase of the singlet $vev$ should be time-dependent, otherwise, such constant phase can be  easily rotated away with the redefining the $V_L$ and $V_R$ ~\cite{McDonald}.
Diagonalizing the quark mass matrix results in some non-diagonal kinetic terms. In addition, couple of time-dependent terms appear in the Lagrangian (see Appendix eq.\ref{D}), namely 
\begin{eqnarray}
 \overline{Q}_L i \gamma^\mu \partial_\mu Q_L+&& \overline{V}_L i \gamma^\mu \partial_\mu V_L\nonumber\\
&& \to \overline{Q'}_L i \gamma^\mu \partial_\mu Q'_L+ \overline{V'}_L i \gamma^\mu \partial_\mu V'_L+ \Delta\mathcal{L}_k + const.\nonumber\\
 \label{potsS}
 \end{eqnarray}
 Since the CP violation disappears for a constant phase, when calculating the baryon asymmetry only the following kinetic term needs to be considered   
\begin{equation}
\Delta\mathcal{L}_k=-{\lambda_V^2 w^2 \over M^2}\dot{\xi}(\overline{Q'}_L \gamma^0 Q'_L-\overline{V'}_L \gamma^0 V'_L).
\end{equation}
Such term increases the baryon density of the universe with the transport of charge into the bubble wall. 
Conventionally, the amount of BAU is calculated via the following relation,
\begin{equation}
n_B = -N_f \int {\Gamma_{sph}(T) \over 2 T} \mu_B dt,
\end{equation}
where $N_f$ is the number of flavors in the model. The sphaleron rate, $\Gamma_{sph}$, is defined as $\Gamma_{sph} = K(\alpha_W T)^4$ in the symmetric phase. $K$ is the numerical factor reflecting the uncertainty in the estimate of the transition rate between vacua of different $B + L$ value, It has been estimated to be between 0.1 and 1 \cite{Ambjorn}. The chemical potential, $\mu_B$, is associated with the baryonic charge in $n_B$. The chemical potential for third generation is as follows ~\cite{McDonald},
\begin{equation}
\mu_{B}=-{5 \over 6}{\lambda_V^2 w^2 \over M^2}\dot{\xi}.
\end{equation}
We assume that the mass parameter $M$ is much larger than the temperature, i.e. $M\geq T$. Thus the sphaleron fluctuations cannot produce V quark pairs and we get the number density of baryons $n_B$ at the temperature $T$ as follows 
\begin{equation}
n_B = {5 K\alpha_W ^4 \over 2 } {\lambda_V^2 w^2 \over M^2 }\delta\xi T^3,
\end{equation}
where $\delta \xi$ is the total change of the phase $\xi$.
The BAU is determined via the ratio of the baryon number to the entropy \cite{{Cline:2006ts}}. The entropy density can be defined as
\begin{equation}
s=\frac{2\pi^2}{45}g^* T^3,
\end{equation}
therefore,
\begin{equation}
{n_B \over s} = {225 K \alpha_W^4\over 4\pi^2 g^*}{\lambda_V^2 w^2 \over M^2 }\delta\xi,
\label{B}
\end{equation}
where the SU(2) gauge coupling $\alpha_W = 3.4\times 10^{-2}$ and $g^*\sim 100$ is the effective number of degrees of freedom in the thermal equilibrium.
The observations from WMAP gives the following value for $n_B / s$ ratio ~\cite{WMAP,Ade:2013zuv}
\begin{equation}
{n_B \over s}= 8.7\pm0.3 \times 10^{-11 }.
\label{N}
\end{equation}
From Eq.(\ref{B}) and Eq.(\ref{N}) we get,
\begin{equation}
K {\lambda_V^2 w^2 \over M^2 }\delta\xi =  1.14\pm0.3 \times 10^{-3}.  
\label{beta}
\end{equation}
The numerical analysis of the equations (\ref{B})and (\ref{N}) have been performed via scanning the involving parameters in the following range,
\begin{eqnarray}
0.3 \; \text{TeV} < &&M < 13 \; \text{TeV} ,\nonumber \\
0<&&\lambda_V <1 ,\nonumber \\
2 \text{GeV} \; \text{GeV}<&&w < 400 \; \text{GeV} ,\nonumber \\
0<&&\delta \xi<\pi,
\label{Bound}
\end{eqnarray}
with numerical factor $K=1$. 
Figure ~\ref{Fig8} illustrates the parameter space allowing the generation of observed BAU for the $n_B/s$ ratio within $2\sigma$. The result were obtained from scanning in the ranges given by Eq.(\ref{Bound}) with the central value of Eq.(\ref{N}). Notice that the range of $w$ is chosen in agreement with our previous analyses \cite{Darvishi:2016gvm}. 
In Fig.\ref{Fig8}(a) the distribution of the parameters $(\delta \xi, w)$ is shown. The parameter space for the $(w, M)$ points is shown in the Fig.\ref{Fig8}(b). Since $M$ and $w$ are independent parameters, their correlation is a direct consequence of the constraint (\ref{beta}). Based on these results, we conclude that our model provide successful BAU.  
\begin{figure*}[ht]
\includegraphics[scale=0.13]{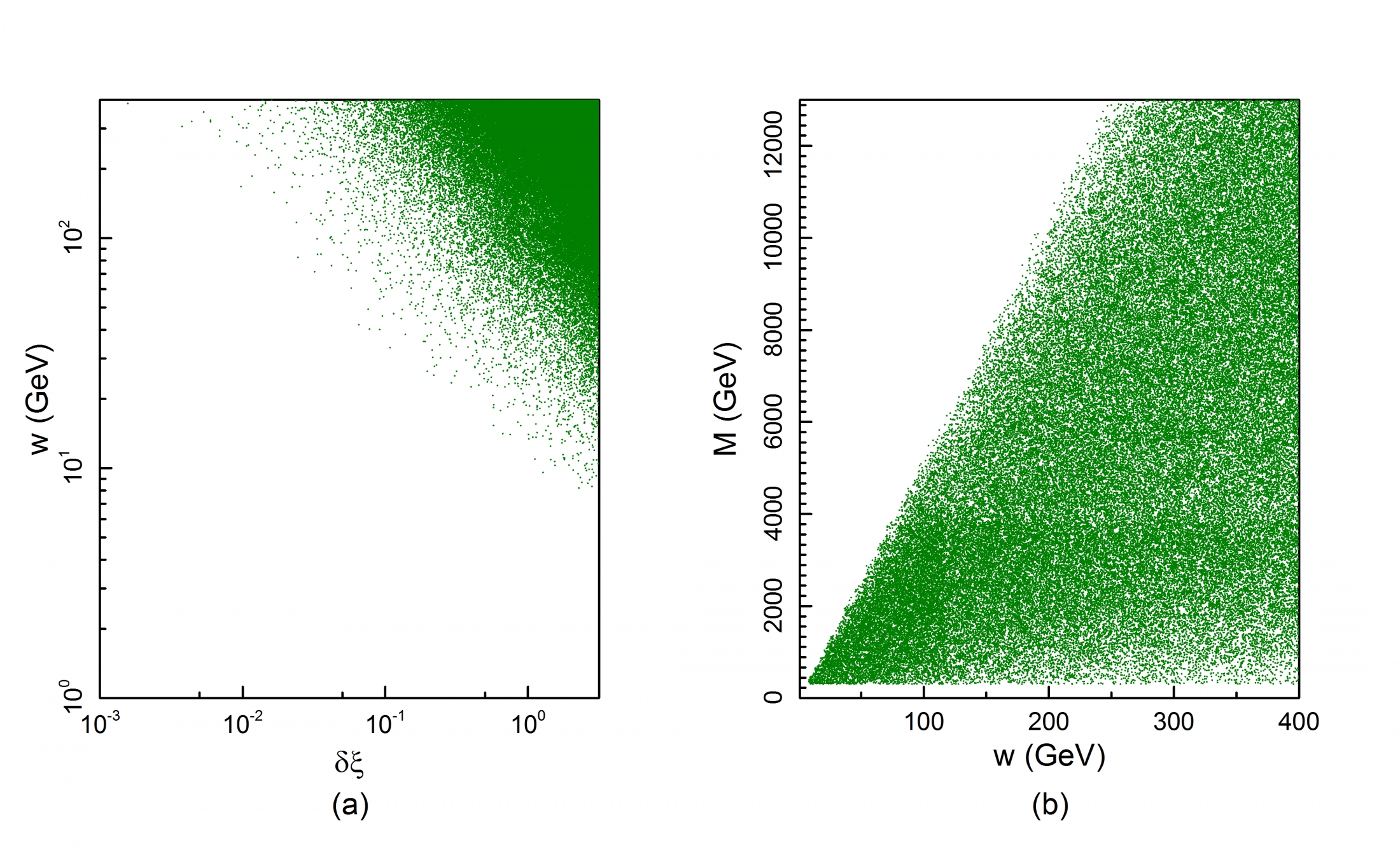}
\caption{The allowed region for $\delta \xi$ , $w$ and $M$ in the Eq.(\ref{N}) of the given ranges, i.e. Eq.(\ref{Bound}), for acceptable BAU within $2\sigma$. (a) is the correlation of $\delta \xi$ and $w$.(b) is the correlation of $w$ versus $M$.}
\label{Fig8}
\end{figure*}
\section{Conclusion}
\label{concl}
In the present work the possibility of the first-order EWPT for the cSMCS model is investigated, showing that such process is strong enough to generate BAU.
In this analysis we have found that cubic terms, $\kappa_2$ , $\kappa_3$ and/or $\kappa_4$ of the potential, should have non-zero values.
This is in agreement with our previous findings, regarding the parameter space of the possible region for CP violation in the cSMCS model, see the reference \cite{Darvishi:2016gvm}. Afterwards, the parameter space of the model for the valid regions of BAU is scanned, concluding that the enlargement of the cSMCS model with a heavy iso-doublet vector quark pair could successfully predict an acceptable value for BAU. 
\begin{acknowledgements}
I am unspeakably grateful to Prof. M. Krawczyk for reviewing this manuscript as well as her discussions and comments.
I also would like to thank M.R. Masouminia and Prof. M. Misiak for helpful discussions. 
This work is supported in part by the National Science Centre, Poland, the HARMONIA project under contract
UMO-2015/18/M/ST2/00518.
\end{acknowledgements}
\section{Appendix}
\subsection{Evaluation of the integral}
\label{A}
The integral Eq.(\ref{I}) is evaluated as follow,
\begin{equation}
{\partial \over \partial y} I_{B,F}(y) = {1 \over 2} \int^\infty_0 dx {x^2 \over (x^2 + y)^{1/2}} 
{1 \over exp((x^2 + y)^{1/2})-1},
\label{A1}
\end{equation}
\begin{eqnarray}
I_{B,F}(y)|_{y=0} =  \int^\infty_0 dx x^2 ln(1-e^{-x})=-{\pi^4 \over 45} ,
\label{A2}
\end{eqnarray}
\begin{eqnarray}
{\partial \over \partial y} I_{B,F}(y)|_{y=0}= {1 \over 2} \int^\infty_0 dx {x \over e^{x}-1}= {\pi^2 \over 12}.
\label{A3}
\end{eqnarray}
\subsection{The field-dependent mass $m_i$}
The field-dependent mass $m_i$ of gauge bosons, Goldestone boson, $m_{\phi_1}$, 
$m_{\phi_2}$ and $m_{\phi_3}$, which is used in Eq.(\ref{thermal}) are given by,
\begin{eqnarray}
{M}_{W}^2   && = {g^2 \phi_1^2\over 4}, {M}_{Z}^2 =( g^2+{g'}^2) {\phi_1^2\over 4},\nonumber \\
{m}_{G}^2   && = \lambda_1 \phi_1^2+\Lambda (\phi_2^2+ \phi_3^2)+ 2\sqrt{2}\kappa_4 \phi_2),\nonumber \\
{m}_{\phi}^2&& = 3\lambda_1 \phi_1^2+\Lambda (\phi_2^2+ \phi_3^2)+ 2\sqrt{2}\kappa_4 \phi_2),\nonumber \\
{m}_{\phi_2}^2 && = 3 \lambda_s \phi_2^2+  \lambda_s \phi_3^2+3\sqrt{2}(\kappa_2+ \kappa_3)\phi_2+{1\over2} \Lambda \phi_1^2,\nonumber \\
{m}_{\phi_3}^2 && = 3 \lambda_s \phi_3^2+  \lambda_s \phi_2^2+\sqrt{2}(-3\kappa_2+ \kappa_3)\phi_2+{1\over2} \Lambda \phi_1^2.\nonumber \\
\end{eqnarray}
$n_i$ is the number of degrees of freedom as,
\begin{eqnarray}
n_W=6, \, n_Z=3, \, n_G=3, \, n_{\phi,\phi_2,\phi_3}=1, \, n_t=12.
\end{eqnarray}
\subsection{Rotation matrix}
As discussed in sec.\ref{baryon} the transformation of $Q$ and $V$ with rotation matrix read $Q^{'}$ and $V^{'}$, shows as follows
\begin{eqnarray}
	\begin{bmatrix}
    Q^{'}_{L} \\
    \\
    V^{'}_{L} \\
	\end{bmatrix}
	&&=
	\begin{bmatrix}
    a & b \\
    \\
    - b^{*} & a^{*} \\
	\end{bmatrix}
	\begin{bmatrix}
    Q_{L} \\
    \\
    V_{L} \\
	\end{bmatrix}	
	\nonumber\\
\alpha &&= \left[ 1 + \left( {\lambda_v w \over M} \right)^2 \right]^{-1/2} \nonumber \\
\beta &&= \left( {\lambda_v w \over M} \right) \left[ 1 + \left( {\lambda_v w \over M} \right)^2 \right]^{-1/2} \;
        e^{-i\xi}
				\label{D}
\end{eqnarray}
\bibliographystyle{JHEP}
\bibliography{bib_CHcol}{}

\bibliographystyle{unsrt}

\end{document}